\documentclass[10pt,letterpaper]{article}
\usepackage[top=0.85in,left=2.75in,footskip=0.75in]{geometry}

\usepackage{amsmath,amssymb}

\usepackage{changepage}

\usepackage{textcomp,marvosym}

\usepackage{cite}

\usepackage{nameref,hyperref}

\usepackage[right]{lineno}

\usepackage[nopatch=eqnum]{microtype}
\DisableLigatures[f]{encoding = *, family = * }

\usepackage[table]{xcolor}

\usepackage{array}

\newcolumntype{+}{!{\vrule width 2pt}}

\newlength\savedwidth

\raggedright
\usepackage[aboveskip=1pt,labelfont=bf,labelsep=period,justification=raggedright,singlelinecheck=off]{caption}

\makeatletter
\renewcommand{\@biblabel}[1]{\quad#1.}
\makeatother

\usepackage{lastpage,fancyhdr,graphicx}
\usepackage{epstopdf}
\fancyheadoffset[L]{2.25in}
\fancyfootoffset[L]{2.25in}
\usepackage{color}
\providecommand{\diff}[1]{{#1}}
\usepackage{physics}

\begin{document}

\vspace*{0.2in}

\begin{flushleft}
{\Large
\textbf\newline{Emergence of economic and social disparities through competitive gift-giving} 
}
\newline
\\
Kenji Itao\textsuperscript{1, 2, 3},
Kunihiko Kaneko\textsuperscript{4, 5*},
\\
\bigskip
\textbf{1} Department of Basic Science, Graduate School of Arts and Sciences, University of Tokyo, Tokyo, Japan.
\\
\textbf{2} Computational Group Dynamics Collaboration Unit, RIKEN Center for Brain Science, Saitama, Japan.
\\
\textbf{3} \textit{BirthRites} Lise Meitner Research Group, Max Planck Institute for Evolutionary Anthropology, Leipzig, Germany.
\\
\textbf{4} The Niels Bohr Institute, University of Copenhagen, Copenhagen, Denmark.
\\
\textbf{5} Research Center for Complex Systems Biology, Universal Biology Institute, University of Tokyo, Tokyo, Japan.
\\
\bigskip

%
%

* kaneko@complex.c.u-tokyo.ac.jp

\end{flushleft}
\section*{Abstract}
Several tiers of social organization with varying economic and social disparities have been observed. However, a quantitative characterization of the types and the causal mechanisms for the transitions have hardly been explained. While anthropologists have emphasized that gift exchange, rather than market exchange, prevails in traditional societies and shapes social relations, few mathematical studies have explored its consequences for social organizations. In this study, we present a simple model of competitive gift-giving that describes how gifts bring goods to the recipient and honor to the donor, and simulate social change. Numerical simulations and an analysis of the corresponding mean-field theory demonstrate the transitions between the following four phases with different distribution shapes of wealth and social reputation: the band, without economic or social disparities; the tribe, with economic but without social disparities; the chiefdom, with both; and the kingdom, with economic disparity and weak social disparity except for an outlier, namely, the ``monarch''. The emergence of strong disparities is characterized by power law distributions and is attributed to the ``rich get richer'' process. In contrast, the absence of such a process leads to exponential distributions due to random fluctuations. The phases depend on the parameters characterizing the frequency and scale of gift interactions. Our findings provide quantitative criteria for classifying social organizations based on economic and social disparities, consistent with both anthropological theory and empirical observations. Thus, we propose empirically measurable explanatory variables and characteristic indices for the evolution of social organizations. The constructive model, guided by social scientific theory, can provide the basic mechanistic explanation of social evolution and integrate theories of the social sciences.

\section*{Author summary}
Human societies have experienced transitions between different types of organizations, including bands, tribes, chiefdoms, and kingdoms. However, quantitative characterizations of the types and mechanisms of transitions have yet to be established. Meanwhile, anthropologists have observed that gift-giving is prevalent in traditional societies and that it enhances social status by imposing reciprocal obligations on others. By constructing a simple model of gift interactions, we demonstrate that competitive gift-giving generates socioeconomic disparities characterized by power law distributions of wealth and social reputation. Consequently, numerical results and corresponding mean-field theory elucidate the above organizational types as distinct phases that depend on the frequency and scale of gift-giving. These findings provide the quantitative criteria for classifying social organizations in human history, together with the potential explanatory variables that can be empirically measured for anthropology, history, and archaeology. This study explains the mechanism of social evolution driven by gift-giving and provides a theoretical framework for the social sciences from the perspective of statistical physics.


\section*{Introduction}
Social scientists have observed that as population or territory size increases, societies are hierarchically organized, and people's roles are differentiated \cite{service1962primitive, turchin2018quantitative}. They have categorized social organizations into several classes, including bands, tribes, chiefdoms, kingdoms, and states \cite{service1962primitive, kang2005examination}. Each class is characterized by different modes of ties (kinship, ideology, or labor division), production systems, and degrees of socioeconomic stratification. 
Ethnographic studies show that the economic disparity is the smallest in hunter-gatherer bands and increasing social disparity with class division characterize chiefdoms and kingdoms \cite{service1962primitive, smith2010wealth, mulder2010pastoralism, smith2010production}.
However, a quantitative characterization of each of the classes and explanations of the transitions between them remain elusive \cite{kang2005examination}. 

To this end, a statistical physics approach based on a simple model is useful in elucidating the existence of qualitatively distinct phases, uncovering the mechanisms by which they emerge, and explaining the reasons why they are universally observed.
Accordingly, we introduce a model based on anthropologists' emphasis on gift-giving as an important form of economic activity in traditional societies (i.e., societies that have not experienced a ``great transformation'' from socially embedded reciprocity to impersonal price-driven market exchanges) and as a driver of social change \cite{mauss1923essai, leach1982social, earle1989evolution, polanyi2002great, graeber2011debt}. Then, we characterize the above social organization classes by their distribution shapes of wealth and reputation and explore their transitions between exponential and power law.

Gift exchange is the prevailing form of economic activity in many traditional societies and plays an important role in shaping economic and social structures.  By analyzing human interactions as exchange processes, several forms of exchanges are observed, including economic exchanges, social exchanges, and ``pure’' gifts. An economic exchange (or market exchange) with currency and pricing markets is a modern form of economic activity \cite{polanyi1957economy} that quantifies the necessary amount of goods to be exchanged \cite{blau1964exchange} and requires certain institutions to enable anonymous exchanges by reducing transaction costs \cite{north2005understanding}. In contrast, social exchanges are conducted based on repetitive interactions and social reputation \cite{mauss1923essai, parsons1956economy, blau1964exchange, polanyi1957economy, north2005understanding, cropanzano2005social}. A social exchange is defined as the voluntary (i.e., uncoerced) act of individuals that is motivated by the reciprocation of what they expect others to return (i.e., delayed reciprocation) \cite{blau1964exchange}. It engenders feelings of personal obligation, gratitude, and trust \cite{blau1964exchange, haas1981trust}. These exchanges are based on the norm of reciprocity, which is considered to be universal \cite{gouldner1960norm}. In contrast, several gifts, namely pure gifts, are donated without seeking reciprocation, especially when such donations are motivated by agapic love \cite{belk1993gift}.

As for the amount of reciprocation, there are many intermediate cases of social exchanges between a pure gift (no reciprocation) and an economic exchange (a complete quantification, sometimes with an interest). By food sharing or charitable gift-giving, donors may acquire a social reputation, but they do not obligate recipients to reciprocate \cite{ready2018wage, bliege2018social, kitanishi1998food, offer1997between, sherry1983gift}. It is also reported to be socially acceptable not to reciprocate when there is a power discrepancy between donors and recipients \cite{lemmergaard2011regarding}.
In contrast, some gifts obligate a reciprocation with an interest, characterized by their competitive nature, known by the expression ``with gifts you make slaves’' \cite{mauss1923essai, blau1964exchange, offer1997between, kelly2013lifeways}. Anthropologists have repeatedly observed such competitive gift-giving in various regions \cite{mauss1923essai, strathern1971rope}.
In many small-scale societies, chiefs give large gifts in public at ceremonies. Thereby, the donors gain prestige, and the recipients are expected to hold ceremonies and make reciprocal gifts larger than the initial gifts within a certain period. If the recipients do not reciprocate or do so too late, they lose their reputation, as observed in the Native American potlatch \cite{mauss1923essai} and the Moka exchange in the Mount Hagen area of central New Guinea \cite{strathern1971rope}. Here, the delay in reciprocation leads to subordination (i.e., social indebtedness), but not to compounding interest as in an economic exchange. Notably, an experimental study revealed that the reciprocation of equivalents increases the social reputation of the donors more than that of the recipients (= reciprocators) \cite{flynn2021better}, which suggests that larger reciprocation is needed to maintain the reciprocators’ reputation.
In this study, we focus on ``competitive gift-giving’' characterized by obligated reciprocation with an interest.

Anthropologists have observed gifts of ornaments and livestock at nuptials, funerals, trades, and various rituals in many regions \cite{strathern1971rope, malinowski1922argonauts}. Furthermore, they discuss the gift as a means of expanding alliances and gaining social status by imposing obligations on others \cite{mauss1923essai, leach1982social, komter2007gifts, moriyama2021gift}. Despite such extensive empirical work, few mathematical studies have been conducted based on these observations, while current mainstream econophysics focuses on market exchanges \cite{yakovenko2009econophysics, chakrabarti2013econophysics}. In contrast, a theoretical analysis of gift-giving will explain how gift interactions shape social organizations in traditional societies.
Previously, we numerically illustrated the evolution of various network structures and the development of economic and social disparities, depending on the frequency and scale of the gift-giving. As gifts are exchanged more frequently and massively, socioeconomic disparities are generated, and social networks become less clustered and more hierarchically organized, which is consistent with ethnographic data \cite{itao2023transition}. However, the numerical model considered both gifts and kinship, which was too complicated to analytically derive the conditions for the transitions in social organizations and socioeconomic disparities.

In this study, we examine the social consequences of competitive gift-giving. We assume that people give, receive, and reciprocate gifts because social norms obligate them. Although consumer researchers have studied the individual strategies involved and have shown that people sometimes refuse to receive gifts \cite{belk1993gift, sherry1983gift}, we consider the ideal case by ignoring such individual strategies. Our goal is to capture the essential mechanisms by which competitive gift-giving drives social change through the consideration of its simplest form. Realistic cases would be considered as deviations from ideal types, reflecting case-specific characteristics \cite{weber1922soziologische}.

The degree of economic and social disparities varies across societies \cite{service1962primitive, smith2010wealth, mulder2010pastoralism, smith2010production}. Both exponential and power law distributions are observed for wealth and social network degree \cite{druagulescu2001exponential, barabasi1999emergence, newman2003social, jusup2022social, schnegg2006reciprocity, schnegg2015reciprocity, yakovenko2009econophysics, chakrabarti2013econophysics}.
Exponential distributions are observed when random interactions and reciprocal relationships are dominant \cite{schnegg2006reciprocity, schnegg2015reciprocity, yakovenko2009econophysics, chakrabarti2013econophysics}. Power law distributions develop when ``rich get richer'' processes, such as preferential attachment, operate \cite{barabasi1999emergence, simon1955class}. In particular, ethnographic research has reported that the degree distribution of the social network follows an exponential distribution when economic inequality is small and a power law when it is large \cite{schnegg2006reciprocity}.
Therefore, in this study, the presence and absence of a strong disparity are characterized by power law and exponential distributions, respectively.

Here, we introduce a simple model for the formation of economic and social structures through gift interactions. Further, we demonstrate the transition in the distribution shapes of wealth and social reputation scores between the exponential and the power law. According to the evolution of distributions, four phases of society are characterized: the band, without economic or social disparities; the tribe, with economic but without social disparities; the chiefdom, with both; and the kingdom, with economic disparity and weak social disparity, except for a ``monarch''. These correspond to the basic types of social organizations in anthropology.
Although the transition in wealth distribution from the exponential to the power law has been discussed in econophysics, we show how gift-giving leads this transition and shapes social organizations. Further, we illustrate the emergence of outliers in the reputation score distribution in the kingdom phase.
Then, we numerically and analytically scrutinize the conditions for the transition.

\section*{Model}
\begin{figure}[tb]
\includegraphics[width=\linewidth]{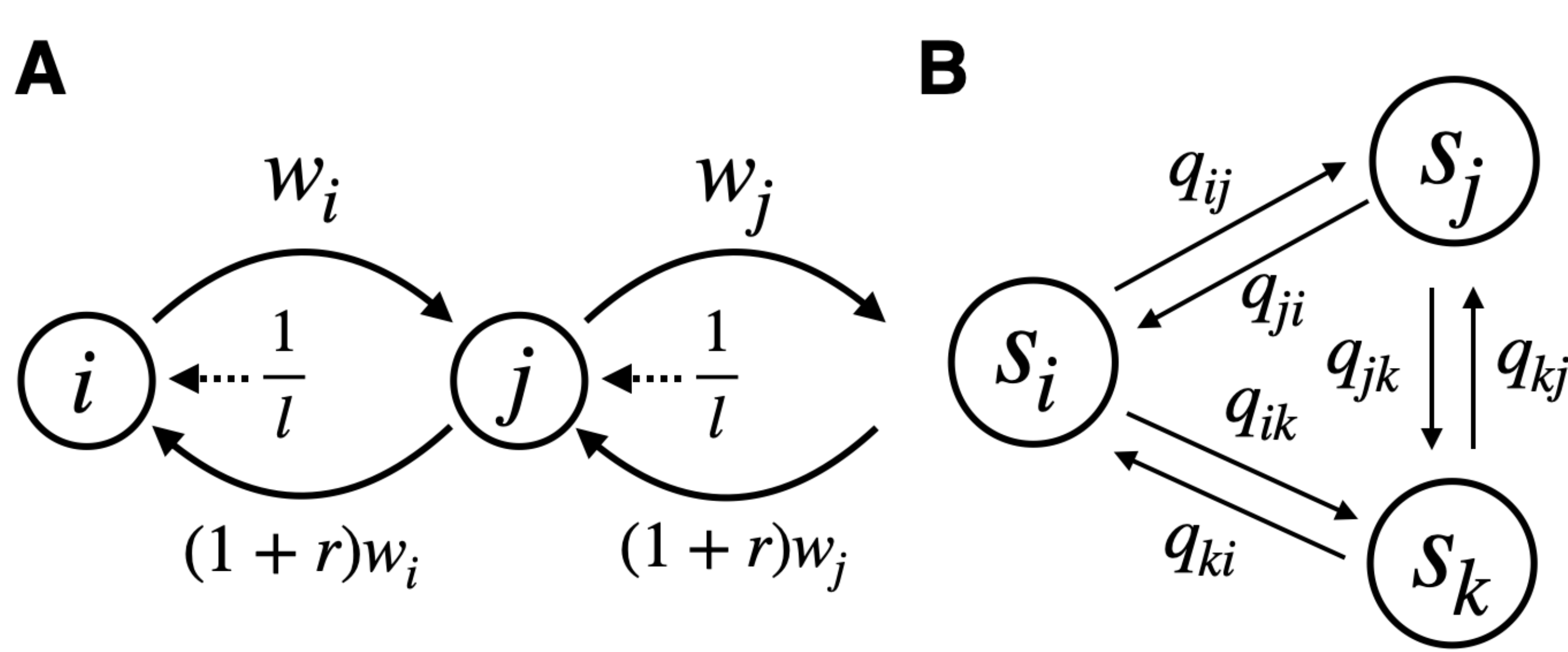}
\caption{\label{fig:gift_model} Schematic of the model. (A) Economic change through gift interaction and production. In each step, individuals give their entire wealth to someone else (top arrows), earn $1 / l$ of wealth (dashed arrows), and then reciprocate for the initial gift with amplification (bottom arrows). (B) Network representation of social relationships. The edge weight $q_{ij}$ represents the number of gift-giving and reciprocation from $i$ to $j$. As wealth is more frequently moved from $i$ to $j$, $j$ becomes more likely to choose $i$ as the recipient in gift-giving. The social reputation score $s_i$, which indicates $i$'s social status, is calculated as the sum of $q_{ik}/l$ for all nodes $k$ (i.e., the out-degree of node $i$).}
\end{figure}

Fig \ref{fig:gift_model} illustrates the schematic of the model. We consider the set of $N$ individuals interacting with each other through gift-giving. In each step, they choose someone else and give their entire wealth $w_i$, following the probability $p_{ij}$ determined by the count of past interactions $q_{ij}$ explained below. Then, they earn $1/l$ of wealth through any form of production and reciprocate for received gifts by amplifying $1 + r$ times. Here, $r$ gives the appropriate interest rate for reciprocation. Note that the reciprocation is delayed due to the intervening production process. When recipients' wealth is insufficient for reciprocation, they will accrue ``debt'' (and donors accrue credit), which will be repaid based on their subsequent production. Debtors cannot bestow gifts before completing repayment. An individual dies with a probability of $1 / l$ and is replaced by a new individual with $w_i = 1 / l$ and $q_{i*} = q_{*i} = 1$. Hence, $l$ gives an expected number of steps each individual experiences before dying, which we term the frequency of gifts in a lifetime. To fix the expected amount of production within a lifetime, the amount of production is scaled as $1/l$. At the time of $i$'s death, debts and credits related to $i$ disappear. 

To model the ethnographic observations that appropriate reciprocation strengthens social ties and that large donations enhance social reputation, we considered the change in social relationships through gift interactions. 
First, we count the number of gift-giving and reciprocation from $i$ to $j$, denoted by $q_{ij}$. Here, we interpret the gift from $i$ to $j$ as ongoing (i.e., $q_{ij}$ increases), while $j$ is indebted to $i$.
The probability that $i$ chooses $j$ as a recipient is given by $p_{ij} = q_{ji} / \sum_k q_{ki}$, i.e., the more $i$ receives from $j$, the more likely $i$ is to give to $j$. We term $s_i = \sum_k q_{ik} / l$ as $i$'s reputation score, which counts the number of gift-giving and reciprocation, indicating social status. In other words, $q_{ij}$ represents the weights of the edge directed from $i$ to $j$, and the reputation scores represent the out-degrees in the social network. 

Hence, our model expresses that proper reciprocity brings equal social ties, whereas failure to reciprocate results in unequal relations. Consider the consequences of a gift from $i$ to $j$. An appropriate reciprocation results in an increase of $q_{ij}$ and $q_{ji}$ by $1$, i.e., the social ties from $i$ to $j$ and from $j$ to $i$ increase equally.
\diff{However, if $i$'s gift is so large that $j$ needs several steps to complete the reciprocation, $q_{ji}$ increases by $1$ (i.e., the reciprocation is counted as $1$), and $q_{ij}$ increases by the number of steps needed for reciprocation (i.e., the gift from $i$ to $j$ is considered ongoing while $j$ is indebted).}
If $j$ is in debt, $j$'s constant production $1/l$ will be paid to $i$, and $i$'s reputation will increase by $1/l$ with each step. Hence, individuals elevate their status by the amount of credit they earn.
\diff{Here, we assume that any form of wealth transfer (gift, reciprocation, and collection of debts) equally strengthens the edge weights. Although there may be differences in the strength of contributions in building social connections, we will show in the Discussion section that considering such differences results in only minor quantitative changes to the following findings.}

In the reciprocation process, the return of the principal amount of the gift received at that step is prioritized. Thus, in the simulation, each individual $i$ receives $w_i$ (the wealth $i$ has given to someone else) simultaneously, and then the interest is repaid in random order. Additionally, if individuals are indebted to multiple donors, the repayment is made to a randomly selected donor in each step. While individuals are in debt, they repay their wealth to the donors at each step (i.e., they do not wait until they can pay the full amount). Thus, indebted individuals do not have wealth at the time of the giving process and consequently do not endow new gifts.

In this model, even though donors give their entire wealth, they receive more (or at least the same) return at the end of the step. Note that if we change this to giving a fraction $p$ of the wealth, the effective interest rate will be $pr$ since donors earn $prw$ of wealth instead of $rw$. Similarly, we can also consider situations in which the donor distributes gifts to multiple individuals by rescaling parameter values. When the donor makes a gift to $k$ people in $k$ equal parts, $l$ increases by $k$ times, and $r$ decreases by $1/k$ times (i.e., $p = 1/k$). To account for such an effective change in the parameters, one need only look at the different parameter regions following the above rescaling. 
Additionally, the amount of production per step can be set to any constant $c$ [wealth/time]. Then, the expected amount of production within a lifetime is $cl$. Now, the gift and return are proportional to wealth, meaning that the result is unchanged by the value of $c$. (One can simply rescale the unit of wealth.) Therefore, without loss of generality, $c$ is set to $1/l$ so that lifetime production is set to $1$. In this $c = 1 / l$, ``$1$'' has the dimension of [wealth], and $l$ has the dimension of [time].

The parameters involved are the interest rate for reciprocation $r$, frequency of gifts in a lifetime (or the life expectancy measured in steps of gift-giving) $l$, and population size $N$, which is set at $N = 100$, unless otherwise mentioned. The initial conditions are set at $w_i = 1 / l$ and $q_{i*} = q_{*i} = 1$ for all individuals.

\begin{figure*}[tb]
\includegraphics[width=\linewidth]{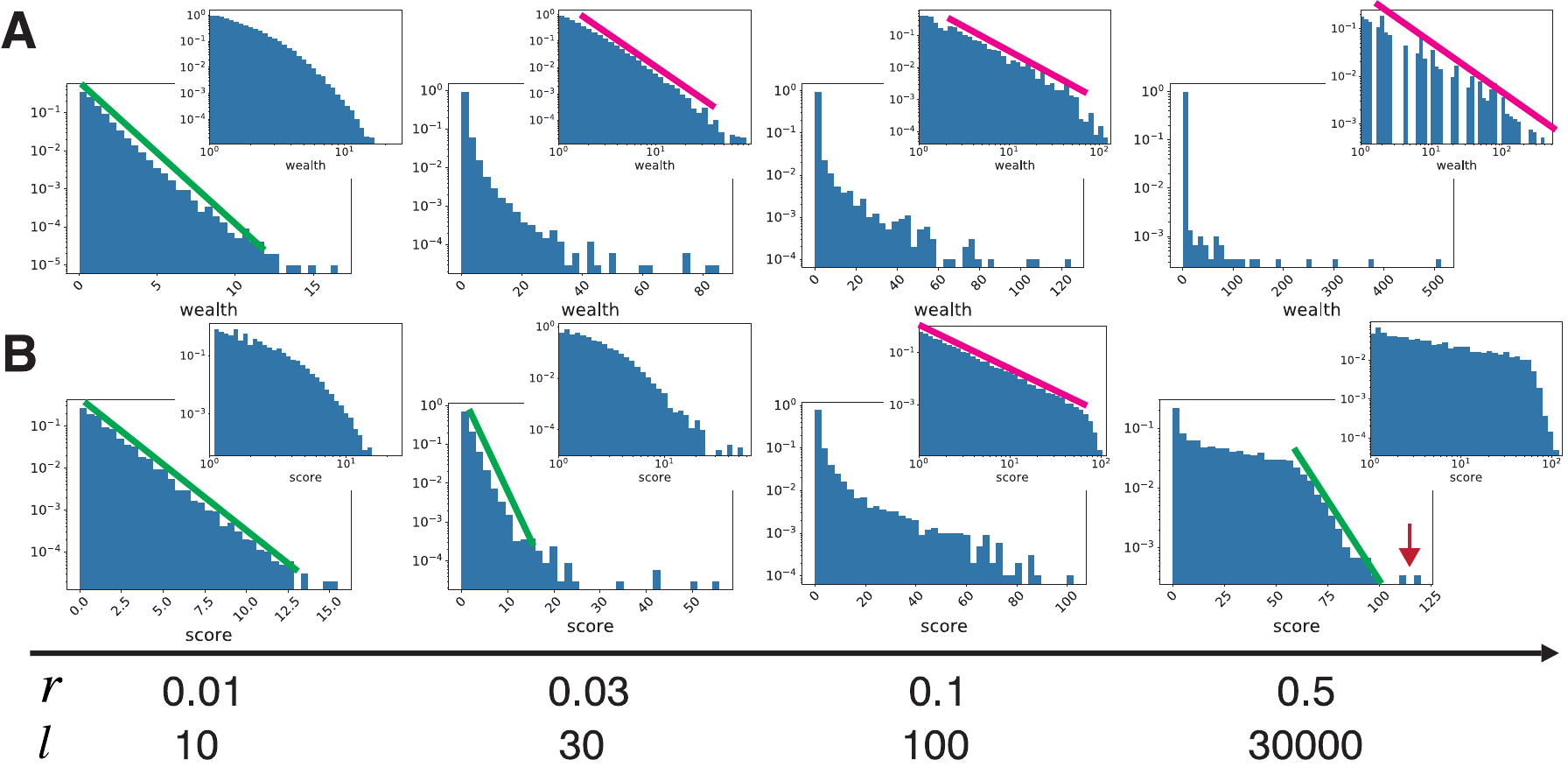}
\caption{\label{fig:gift_distribution} Distributions of wealth (A) and reputation score (B) for different values of the interest rate $r$ and frequency of gifts $l$. We recorded the wealth and reputation scores of people at their deaths in the last $9 \times 10^6$ steps to describe their stationary distributions. Graphs show the semi-log and log-log plots of the distributions. For each distribution, the top 30\% are fit via both exponential and power law distributions, and the one with the smaller error is plotted. The downward arrow in the bottom right graph represents monarchs.}
\end{figure*}

\begin{figure*}[t]
\includegraphics[width=\linewidth]{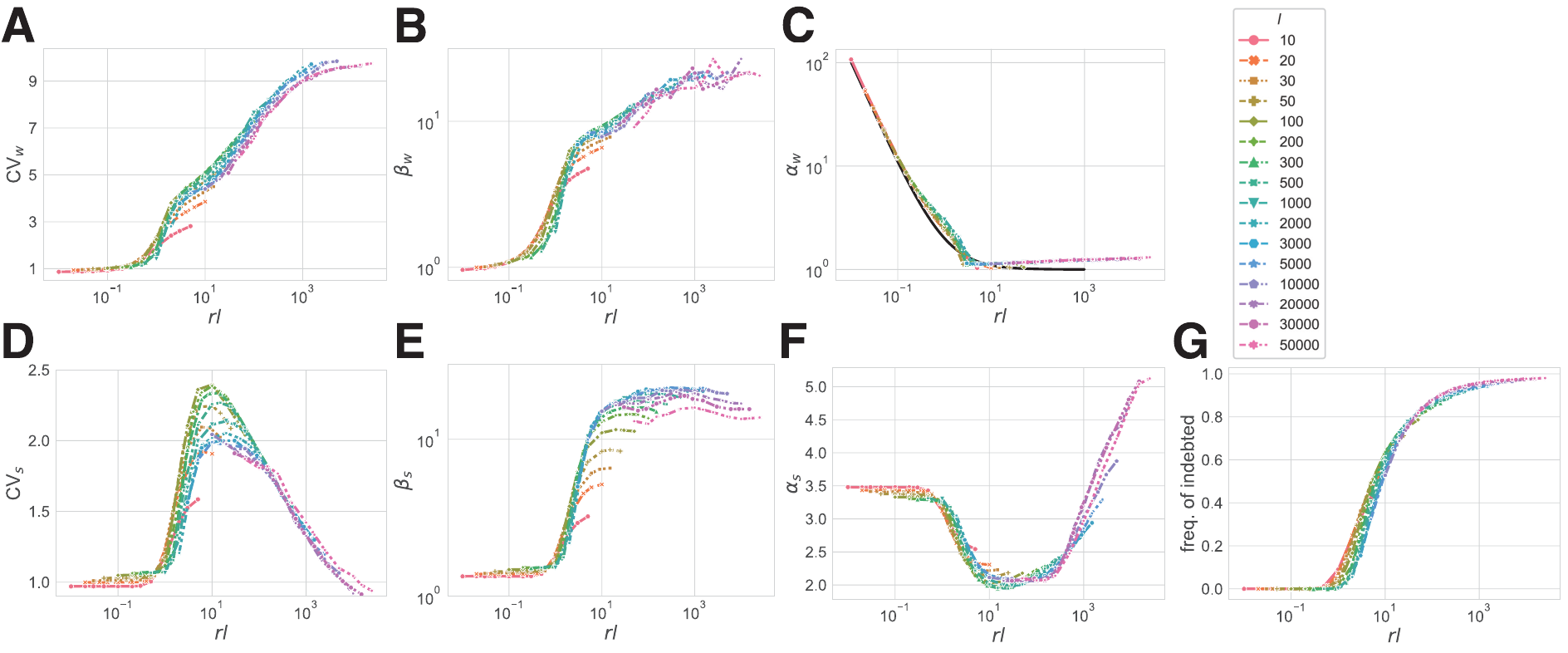}
\caption{\label{fig:gift_indices} Statistical characteristics of the simulation results. (A, D) Coefficients of variation $CV$ (SD/mean) for the wealth and reputation score distributions, which equal 1.0 for exponential distributions and are greater for power law distributions. (B, E) The half-lives of the wealth and score distributions obtained via exponential fitting $\exp(- x / \beta)$. (C, F) Power exponent of the wealth and reputation score distributions resulting from power law fitting $x^{-\alpha}$. The black line shows the theoretical estimates $\alpha_w = 1 + 1 / rl$. (G) The frequency of debtors in society.
The different colors of the lines represent the values of $l$ (as shown in the legend in the right box).}
\end{figure*}

\begin{figure}[t]
\includegraphics[width=\linewidth]{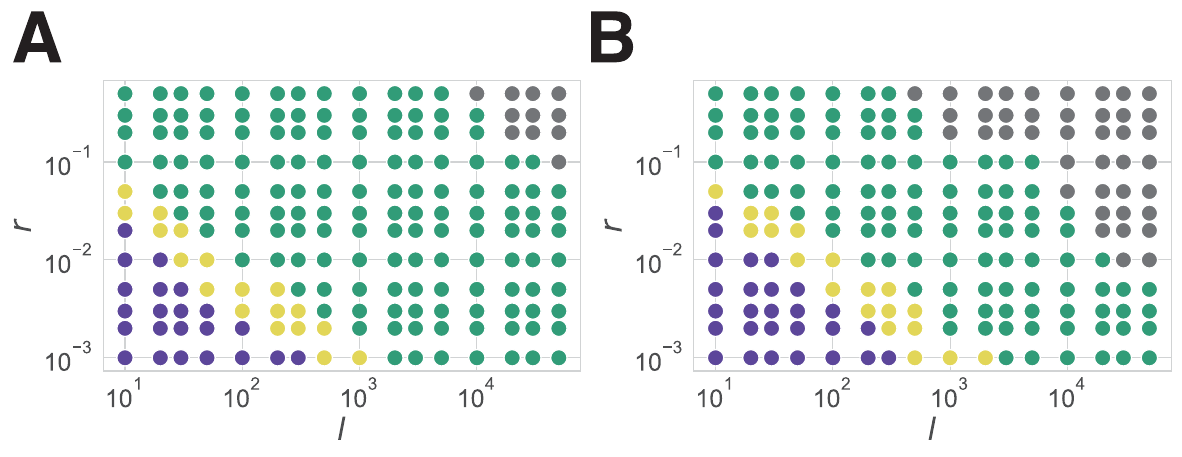}
\caption{\label{fig:gift_phase} 
Phase diagram of social organizations against the interest rate $r$ and frequency of gifts $l$. (A) The figure shows the parameter regions for the band phase $CV_w, CV_s < 1.1$ (purple); tribe phase $CV_w \ge 1.1$ and $CV_s < 1.1$ with $\beta_s < 10$ (yellow); chiefdom phase $CV_w, CV_s \ge 1.1$ (green); and kingdom phase $CV_w \ge 1.1$ and $CV_s < 1.1$ with $\beta_s \ge 10$ and a monarch (grey). (B) Diagram using the relative error of the fittings $RE$, given by $\log$(the error of the exponential fit/the error of the power-law fit).
The figure shows the parameter regions for $RE_w, RE_s < 0$ (purple); $RE_w \ge 0$ and $RE_s < 0$ with $\beta_s < 10$ (yellow); $RE_w, RE_s \ge 0$ (green); $RE_w \ge 0$ and $RE_s < 0$ with $\beta_s \ge 10$ and a monarch (grey).
}
\end{figure}

\section*{Results}
\subsection*{Numerical results}
Simulations were performed for $10^7$ time steps, within which stationary distributions of wealth and reputation score were realized. Fig \ref{fig:gift_distribution} shows the distributions of wealth and reputation score at the time of death in the last $9 \times 10^6$ steps, which characterizes the emergent social organizations.

Distributions are fit by either an exponential or a power law distribution. It illustrates that as the interest rate $r$ and the frequency of gifts $l$ increase, the wealth distribution shifts from exponential to power law first, and then the reputation score distribution follows. However, the reputation score reverts to an exponential distribution for extremely large values of $r$ and $l$. A large disparity in reputation scores indicates a strong network bias. Hence, it is difficult for newcomers to establish relationships with influential people when $r$ and $l$ are large.

We performed the simulation $100$ times for each condition. Fig \ref{fig:gift_indices} presents the average statistical characteristics of the simulation results. We estimated the transition in wealth and score distributions using the coefficients of variation $CV$ (standard deviation/mean) shown in Fig \ref{fig:gift_indices}A and D, which equal 1.0 for an exponential distribution and are greater for a power law distribution.
Fitting gives the half-lives of the exponential distribution $\beta$ and the power law exponent $\alpha$ (Fig \ref{fig:gift_indices}B, C, E, and F). Fig \ref{fig:gift_indices}G shows the frequency of debtors in society.
Fig \ref{fig:gift_indices} suggests that these depend only on $rl$. This $rl$ roughly gives the growth rate of wealth per generation, as individuals receive the interest of $r$ a total of $l$ times.
As $rl$ values increase, the wealth and score distributions successively shift to power law distributions, and then the score distribution reverts to an exponential distribution with a longer half-life than that observed for small $rl$ values.

\begin{figure*}[tb]
\includegraphics[width=\linewidth]{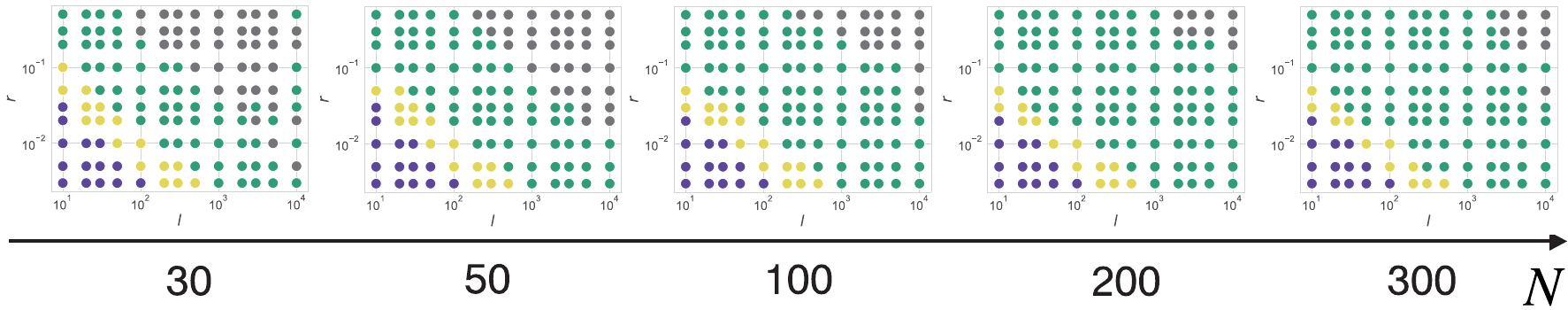}
\caption{\label{fig:gift_phase_N} Dependence of phase diagram on the population size $N$. This figure shows the phase diagrams of social organizations against the interest rate $r$ and frequency of gifts $l$, whose boundaries are identified based on $RE$s, as in Fig \ref{fig:gift_phase}B.
}
\end{figure*}

Fig \ref{fig:gift_phase}A displays the phase diagram of the emergent social organizations against $r$ and $l$. The phases sequentially shift from ``band'' to ``tribe'' at $rl \simeq 0.25$, to ``chiefdom'' at $rl \simeq 1.0$, and eventually to ``kingdom'' at $rl \simeq 3000$. Here, phases are identified via $CV$ values. 
Fig \ref{fig:gift_phase}B indicates that the phase diagram remains almost unchanged (except for the kingdom phase), even when phases are identified via the relative error of the exponential and power law fittings.
Moreover, Fig \ref{fig:gift_phase_N} shows that the phase boundaries at $rl \simeq 0.25$ and $1$ are independent of the population size $N$, whereas the condition for the emergence of the kingdom phase depends on $N$.

\begin{figure*}[tb]
\includegraphics[width=\linewidth]{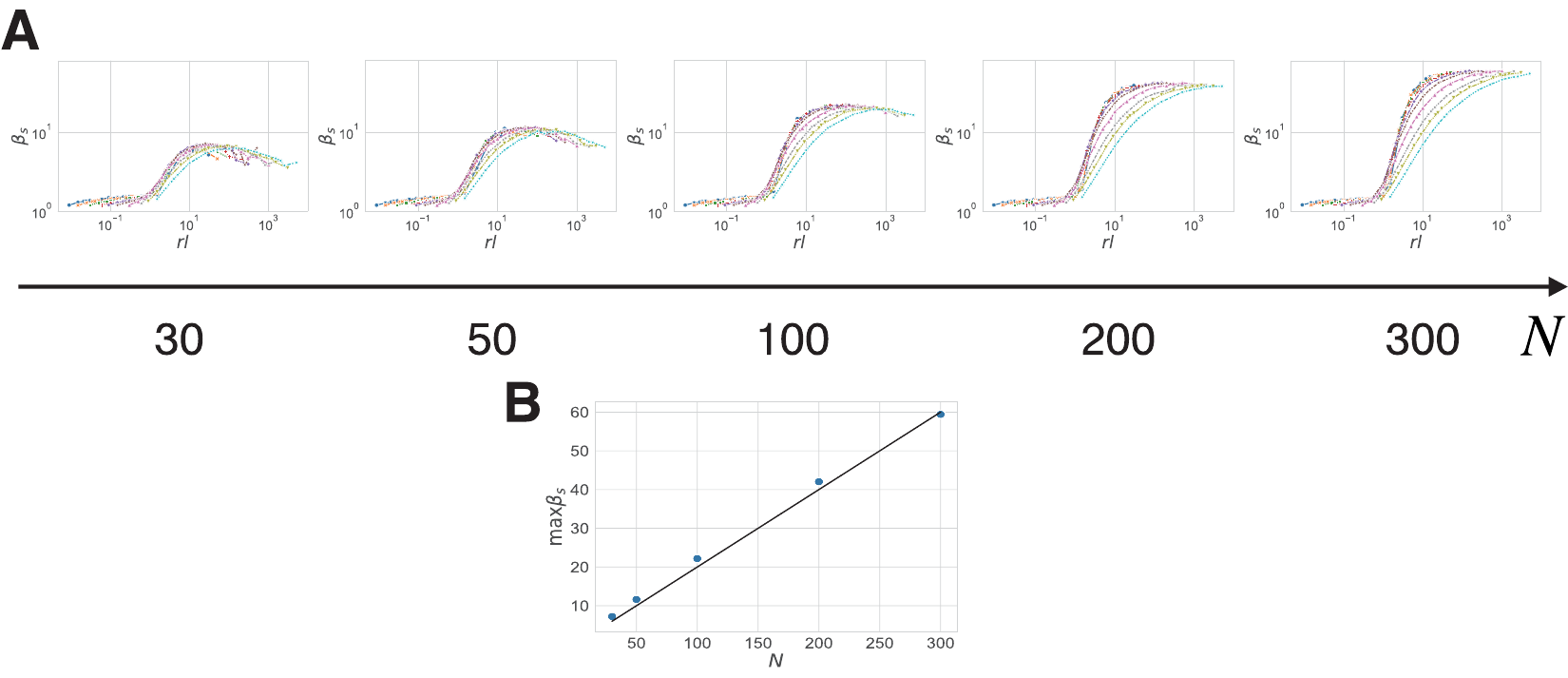}
\caption{\label{fig:gift_beta_s} Dependence of the half-lives of the score distribution $\beta_s$ on the population size $N$. (A) Dependence of $\beta_s$ on $rl$ for different population sizes $N$. (B) The relationship between $N$ and the maximum values of $\beta_s$ when changing $r$ and $l$. The black solid line is the regression line $\max_{r, l} \beta_s = 0.2 N$.
}
\end{figure*}

The half-life of score distribution $\beta_s$ increases with $rl$ and converges to certain values depending on $N$ (Fig \ref{fig:gift_indices}E). 
Fig \ref{fig:gift_beta_s} illustrates the simulation results for different population sizes $N$ and suggests that the asymptotic values of $\beta_s$ are around $0.2 N$.

The observed trend can be roughly explained as follows: Since gift interactions involve amplified reciprocation, economic disparity intensifies with $rl$. As the economic disparity becomes sufficiently large, many individuals fail to reciprocate. Thus, social disparity is exacerbated. Hence, stronger disparities emerge for larger values of $rl$. When economic disparity is extreme, an opulent monarch appears. The other people earn reputation scores only before they become debtors to the monarch, resulting in the social disparities among the vast majority diminishing.

\subsection*{Analytical estimation for the transitions in economic disparity}
To analytically estimate the aforementioned socioeconomic change, we first consider the economic change that occurs when $rl$ is small. In this regime, the score distribution is exponential, and the preferential attachment for recipient choice is weak. Therefore, in our analysis of economic change, we make the simplifying assumption that recipients are chosen with uniform probability. This assumption is further verified in the following analysis of social change.

Wealth changes via production and reciprocation. At each step, the wealth of $i$ increases with production by $1/l$ and with being reciprocated by $rw_i$, whereas it decreases with reciprocating by $rw_j$. 
Therefore, by replacing the difference in wealth of adjacent time steps with $\dot{w}$ and coarse-graining the discrete time evolution to be continuous, the wealth dynamics are given by
\begin{align}
    \dot{w_i} &= 1 / l + r \left(w_i - \sum_j w_j \eta_j \right) =: f(w_i); \label{eq:langevin}\\
    &= 1 / l + r \left(w_i - \ev{w}\right) - r \left(\sum w\eta - \ev{w}\right), \label{eq:langevin2}
\end{align}
where $\ev{x}$ represents the average value of $x$ across the society. This formulation assumes that reciprocation is appropriately performed.
Here, $\eta$ is a random variable equal to $1$ with probability $1 / N$ and $0$ otherwise, and its time correlation is negligible under small network bias.
Eq. \eqref{eq:langevin2} clarifies three terms indicating the stable growth of wealth owing to production, exponential growth owing to received interest, and stochastic fluctuation occasioned by the random choice of recipient, respectively. When $i$ receives too many gifts and cannot reciprocate, $f(w_i) < -w_i.$ At that time, $i$ will have a debt of $f(w_i) - w_i$, and $w_i$ becomes $0$, i.e., $w$ is always nonnegative. Note that $\ev{\dot{w}} = 1 / l$ and the life expectancy is $l$, leading to $\ev{w} = 1.$

Then, we calculate the temporal development of the wealth distribution $P(w)$. Now, each individual's wealth grows at a rate of $\dot{w}$ and disappears with a probability of $1 / l$.
The term $\sum w\eta$ in $\dot{w}$, representing the amount of interest that $i$ should pay for reciprocation, fluctuates randomly depending on the number and nature of individuals that endowed gifts to $i$. First, let us ignore this fluctuation and assume $\sum w\eta = \ev{w}$. Then, by neglecting the last term in Eq. \eqref{eq:langevin2}, the temporal development of the wealth distribution is given by 
\begin{equation}
    \frac{\partial P}{\partial t} = - \frac{P}{l}  -\frac{\partial }{\partial w}\left(\frac{1}{l} + rw - r\ev{w}\right)P.\ (\text{for }w \neq 1/l) \label{eq:FP}
\end{equation}
Its steady-state condition satisfies
\begin{equation}
    0 = - P 
-\frac{\partial P}{\partial w} - rl\frac{\partial }{\partial w}\left(w - \ev{w}\right)P.\ (\text{for }w \neq 1/l) \label{eq:FP2}
\end{equation}
For $w = 1/l$, there is an inflow of $1/l$ due to new births replacing the dead.
Although the integral of the probability function decreases in the regions with $w \neq 1/l$, this inflow at $w = 1 / l$ compensates the decay so that $\int P dw = 1$ is conserved.
This inflow ensures that $\ev{w} = 1.$ Then, by solving Eq. \eqref{eq:FP2}, noting that $\ev{w} = 1$ and $\int P(w)dw = 1$, we obtain
\begin{equation}
     P_1(w) = \frac{(1 / rl - 1)^{1 / rl}/rl}{(w - 1 + 1/rl)^{1 + \frac{1}{rl}}}.\ \ 
(w \neq 1 - 1 / rl)
\end{equation}
The solution $P_1(w)$ is self-consistent because it satisfies $\int wP_1(w)dw = 1$.
If $rl < 1$, then $P_1 (w)$ is valid for $w > 0$.
Otherwise, $P_1(w)$ diverges at $w = 1 - 1 / rl$. However, in reality, there is no divergence because the fluctuation term that we ignored in Eq. \eqref{eq:langevin2} appears as the diffusion term in Eq. \eqref{eq:FP}. Additionally, individuals with $w < 1 - 1 / rl\Leftrightarrow 1 / l + rw < r \ev{w}$ lose their wealth by reciprocating and eventually accrue debts. Then, $w$ remains at $0$ until repayment is completed.
Therefore, the wealth distribution for the rich follows $P_1(w)$, whereas that for the poor peaks at $w = 0$. The power exponent $\alpha_w$ equals $1 + 1 /rl$, indicating a greater disparity for larger values of $rl$. Note that the condition for frequent incurrence of debts is given by $rl > 1$, which is consistent with Fig \ref{fig:gift_indices}G.

In contrast, when the growth term $r(w - \ev{w})$ is negligible, the wealth change is dominated by random exchanges, denoted by the last term in Eq. \eqref{eq:langevin2}. Noting that random exchanges of energy lead to Boltzmann distribution if the energy is conserved and the exchanges have finite variance, as discussed in econophysics literature \cite{yakovenko2009econophysics, chakrabarti2013econophysics}, the steady distribution follows $P_2(w) = \exp(- w)$ because $\ev{w} = 1$.

We now estimate the condition for the transition from the exponential distribution $P_2$ to the power law distribution $P_1$. The power law dominates when the second (growth) term surpasses the last (fluctuation) term in Eq. \eqref{eq:langevin2}. Let us consider the wealth dynamics of rich individuals. 
When $w_i - \ev{w}$ is typically larger than $\sum w\eta - \ev{w}$, their wealth grows exponentially, and the power law distribution $P_1$ is therefore obtained. 
Thus, the deviation of the rich must surpass the variance anticipated by the Boltzmann distribution.
The variance of the growth term $w - \ev{w}$ equals the variance of $P_1(w)$, that is, $1 / (1 - 2 rl)$.
In contrast, that of the fluctuation $\sum w\eta - \ev{w}$ equals the sum of the variance of $P_2(w)$ and that derived by sampling from the binomial distribution, that is, $1 + (1 - 1 /N)$.
Hence, the power law distribution is obtained if
$1 / (1 - 2 rl) > 2 - 1 /N$, that is, if $rl \gtrsim 0.25$.
Note that the transition at $rl = 0.25$, power law exponent $\alpha_w = 1 + 1 / rl$ for $rl > 0.25$, and half-life $\beta_w = 1$ for $rl < 0.25$ are consistent with the results shown in Figs \ref{fig:gift_indices} and \ref{fig:gift_phase}.

\subsection*{Analytical estimation for the transitions in social disparity}
Next, we analyze the change in the distribution of the reputation score $P(s)$.
Scores increase when individuals receive gifts, reciprocation, or repayment. One acquires credit when the interest $rw_i$ exceeds the recipient's wealth $w_j$. Debtors repay $1 / l$ of wealth (i.e., production), and creditors acquire $1 / l$ of score in each step. Thus, at the completion of the repayment, the creditors acquire scores equal to the amount of credits acquired, whose amount is given by $|rw_i - w_j|_+ (:= \max(0, rw_i - w_j))$. However, because $j$ may die before completing the repayment and $j$'s life expectancy is $l$, the expected increase in the score from the interaction is $\min(|rw_i - w_j|_+, 1)$. Therefore, the temporal development of the score is given by 
\begin{equation}
    \dot{s_i} = 1 / l + \sum \eta / l + \min(|rw_i - w_j|_+, 1). \label{eq:score_langevin}
\end{equation}
Recall that we have already demonstrated that economic disparity, denoted by $\alpha_w$, increases with $rl$ and that $rl > 1$ gives a condition for the frequent incurrence of debts.
Hence, with $rl$, $|rw_i - w_j|_+$ increases, and the score distributions change.

When $rl < 1$, most gifts are reciprocated immediately. In other words, $rw_i - w_j$ is generally negative, and the third term in Eq. \eqref{eq:score_langevin} is negligible. Then, $\ev{\dot{s}} = 2 / l$ and $s_i = 2t_i / l,$ where $t_i$ is $i$'s age. Since the age distribution is $P(t) = \exp(-t / l) / l$, the score distribution is $P(s) = \exp(-s / 2) / 2$. 

When $rl > 1$, some gifts are not reciprocated and debts are incurred. Hence, $\min(|rw_i - w_j|_+, 1) = rw_i - w_j$ as long as the debt is repayable before death. Note that individuals acquire credit only if they give to those who have insufficient wealth. By denoting its probability as $p$ and averaging the stochastic term, we obtain the following estimate:
\begin{align}
    \ev{\dot{s_i}} &= 2 / l + p|rw_i - \ev{w}|_+,  \\
        &= 2 / l +  p|\dot{w_i} - 1 + r - 1 / l|_+.
\end{align}
Hence, $\dot{s}\to p\dot{w}$ if $\dot{w} \gg 1$. Since wealth distribution follows a power law when $rl > 1$, the score distribution also follows a power law.
However, for small $\dot{w} ( < 2 / l)$, $\dot{s}\to 2 / l > \dot{w}$ .
Therefore, the score distribution has more middle classes than does the wealth distribution, resulting in $\alpha_s > \alpha_w$ (see Figs \ref{fig:gift_distribution} and \ref{fig:gift_indices}). The transition of the score distribution from exponential to power law at $rl \simeq 1$ is consistent with the simulation results in Fig \ref{fig:gift_indices}. Hence, the score distribution is exponential at $rl \simeq 0.25$, which validates the ignorance of connection bias in the analysis of economic change, down to such $rl$ values.

When $rl$ is sufficiently large, debtors often die before completing repayment. Hence, the expected increment of score from credit is generally $\min(rw_i - w_j, 1) = 1$. Thus, $\ev{\dot{s}} = 2 / l + p$, leading to
\begin{align}
    s_i &= pt_i + 2t_i / l\simeq pt_i.
\end{align}
Then, the score distribution reverts to exponential if the distribution of $t$ is exponential. Here, $t$ indicates the number of gifts that one makes, which is generally different from the age, in contrast to cases in which $rl < 1$.
If individuals receive gifts from the rich, they will need many time steps for repayment and will no longer earn scores. (Note that once individuals are debtors, their social connections to the rich are strengthened. Since rich people can reciprocate gifts, they will have little chance of earning credits even after completing repayment.)
Hence, $t$ generally equals the time that passes before an individual receives a gift from the richest individual, whom we term the ``monarch''. As the monarch chooses recipients randomly from $N$ individuals, the distribution of $t$ is determined by $P(t) = \exp(-t / N) / N$. Subsequently, the score distribution, except for the monarch, is $P(s) = \exp(-s / pN) / pN$. On the other hand, the expected value of $t$ for the monarch is the life expectancy $l$. Hence, the score distribution will be exponential with a half-life proportional to $N$ for most individuals and $l$ for the monarch. 
Still, the gifts from ordinary people may be repayable, and the score distribution for the middle class will still follow the power law because $\min(|rw_i - w_j|_+, 1) = rw_i - w_j$. Additionally, the monarch cannot find debtors if all others are already his or her debtors. Therefore, the largest score is bounded by $\mathcal{O}(N)$. Because the kingdom phase significantly depends on the interaction between the monarch and the rest of the population, the condition for its emergence depends on the population size $N$, as shown in Fig \ref{fig:gift_phase_N}.
The numerical results for the different values of $N$ in Fig \ref{fig:gift_beta_s} show that $\beta_s$ increases with $rl$ and converges to $0.2N$, indicating that $p \simeq 0.2$.
This value is reasonable, as preferential attachment limits the possibility of acquiring credit by giving gifts to previously untraded individuals. Note that credit is acquired only when the recipient is not indebted to the donor.

To summarize, as $rl$ increases, the score distribution shifts from an exponential distribution with a half-life of $\beta_s = 2$ to a power law, and it then reverts to an exponential distribution with $\beta_s \simeq 0.2 N$, which is consistent with the results shown in Fig \ref{fig:gift_indices}. The transition to a power law at $rl = 1$ is consistent with the results shown in Fig \ref{fig:gift_phase}. Again, a power law emerges by reflecting the ``rich get richer'' process in wealth, and its absence due to either weak economic disparity or suppression by the monarch leads to an exponential distribution.

\section*{Discussion}
In this study, we built a model of competitive gift-giving and numerically demonstrated the emergence of several phases of social organizations that are quantitatively characterized by the shape of the wealth and reputation score distributions, i.e., (i) the band phase, in which both distributions are exponential; (ii) the tribe phase, in which only the wealth distribution obeys a power law; (iii) the chiefdom phase, in which both distributions are power laws; and (iv) the kingdom phase, in which the score distribution is exponential for all individuals except the monarch. This result provides theoretical support for the empirical findings that band societies have less economic inequality than do the other classes, that the presence of social inequality distinguishes chiefdoms and kingdoms from the others, and that monarchs are outliers in the distribution \cite{service1962primitive, kang2005examination, smith2010production, sornette2009dragon}.
Then, we analytically explained their transitions, whose boundaries are defined by $rl$, which is the product of the interest rate and frequency of gifts. 

In addition to the degree of socioeconomic disparities, anthropologists characterized the band as a kinship-based society, the tribe as a larger unit knit by ideological ties such as siblinghood, the chiefdom as a hierarchical organization with class division, and the kingdom as a hierarchical organization with stable royal families \cite{service1962primitive, kang2005examination}. 
Combined with our previous research \cite{itao2023transition}, we can characterize each phase with salient network structures, i.e., the band phase with small kin clusters, the tribe phase with large clusters, and the chiefdom and kingdom phases with a network hierarchy. For such a characterization, it is essential to elucidate these classes as distinct phases. Thus, this study provides a theoretical basis for discussing human social organizations.

Social scientists have argued that increased population density and surplus production will accelerate people's interactions, including gifts, and promote social stratification \cite{service1962primitive, bataille1949part, von2019dynamics}. However, we acknowledge that the quantitative data are inadequate compared with the data of the modern market economy in economics and econophysics.
Still, we previously estimated the gift degree, which reflects the scale and frequency of gifts $r$ and $l$, by using the Standard Cross-Cultural Sample ethnographic database and analyzing the scale of gifts exchanged in marriage celebrations as well as the frequency of gift transactions in response to compensation demands, among other data. We also measured the degree of economic and social inequality by analyzing the numbers of rich, poor, and dispossessed people as well as the degree of social stratification, and we confirmed that economic inequality appears first and social inequality appears second as the gift degree increases \cite{murdock1969standard, kirby2016d, itao2023transition}. Furthermore, our parameter values of the interest rate $r$ and the frequency of gifts $l$ can be estimated directly from ethnographic records of gift ceremonies (e.g., \cite{strathern1971rope}) and from archaeological records of burials (cf. \cite{tsujita2006formation, harke2014grave}).

Here, we propose the interest rate $r$ and the frequency of gifts $l$ as basic explanatory variables to be measured for discussing social organizations. The value of $l$ depends on the types and situations of the gifts. For example, the opportunity for gifts at marriage is proportional to the number of family members and is on the order of $1-10$. In contrast, gifts of livestock and jewelry at ceremonies occur at a much higher frequency \cite{mauss1923essai,strathern1971rope}. 
In the Moka exchange in Hargeners in Papua New Guinea, $r \simeq 1$, meaning that reciprocation should be doubled \cite{strathern1971rope}, whereas in the Kula exchange in the Trobriand Islands, $r \simeq 0$, meaning that equivalents are exchanged \cite{malinowski1922argonauts}.
Ethnographic reports suggest that the increase in tradable goods leads to large values of $r$ and $l$ \cite{mauss1923essai,strathern1971rope}.
Moreover, our results suggest that the study of the shapes of the distributions of goods and reputation in ethnography, archaeology, and historical studies will help to establish the quantitative classification of social organizations. (Recall the ethnographic reports showing that the degree distribution of the social network is an exponential or a power law, depending on the degree of economic inequality \cite{schnegg2006reciprocity}).

Exponential and power-law distributions have been observed in many social systems \cite{yakovenko2009econophysics, chakrabarti2013econophysics, cho2014physicists, tao2019exponential, barabasi1999emergence, newman2003social}. Generally, the power laws originate from a ``rich get richer'' process, whereas exponential distributions arise from random exchanges or reciprocity \cite{cho2014physicists, schnegg2006reciprocity}. In this study, we demonstrated the transitions of the distribution shapes of wealth and reputation scores through gift interactions.
Polanyi proposed reciprocity, centralized redistribution, and market exchange as basic modes of economic activity and stressed that economic activities are inseparable from political and social interactions \cite{polanyi1957economy}. Such interrelationships have been dismissed in present mainstream econophysics, which focuses on market exchanges \cite{yakovenko2009econophysics, chakrabarti2013econophysics}.
However, in our model, reciprocity works in the band and tribe phases because most gifts are reciprocated appropriately, whereas centralized redistribution emerges in the chiefdom and kingdom phases as a constant flow of repayment from the vast majority to rich individuals. 

The dynamics in the ``kingdom phase'' resemble the merge-and-create process, where $N$ variables are set and two randomly chosen values are replaced by their sum and $1$ \cite{takayasu1989steady, minnhagen2004self}. The largest value among the variables then increases infinitely with time, and the remaining variables exhibit a power law distribution. As in our model, the richest exceptional outlier follows distinct dynamics and is sometimes termed the ``dragon-king'' \cite{laherrere1998stretched, sornette2009dragon, sornette2012dragon}. In our model, such an outlier suppresses the activities of others, just as ``kings (monarchs)'' exclude the majority from profiting from the economy across history \cite{robinson2012nations}.

Our model assumes that people always endow gifts and (try to) appropriately reciprocate. However, people may strategically choose different behaviors, which should be further explored \cite{sherry1983gift, belk1993gift}. Intuitively, one could benefit from accepting a gift and not reciprocating, although such a betrayal without reciprocation would damage one's reputation. Then, one would not necessarily endow a gift at the risk of such betrayal. To discuss the evolution of gift-giving with the obligations to give and to reciprocate, it is necessary to consider the process by which people come to take appropriate actions to avoid being ostracized, even when doing so is economically costly. In this regard, studies on altruistic reciprocity and signaling theory on a high-cost display to enhance reputation are instructive \cite{fehr1998gift, hetzer2013co, bliege2018social}.

In the context of social exchange theory, not all transactions are competitive \cite{blau1964exchange, moriyama2021gift}. Indeed, food sharing is done in many hunter-gatherer societies to seek acknowledgement without expecting reciprocation \cite{kitanishi1998food, ready2018wage, bliege2018social}. Such situations will correspond to the case with $r < 0$. Then, the current mechanism does not function to generate economic disparity. However, if economic disparity is generated by other mechanisms, the current mechanism can explain the improvement in status through the distribution of wealth, i.e., the embedding of economic disparities in social disparities. Notably, Malinowski has noted that ``(in the Trobriands) the main symptom of being powerful is to be wealthy, and of wealth is to be generous'' \cite{malinowski1922argonauts}.

\diff{We assume that in gift-giving, each individual gives their entire wealth to one recipient. However, this assumption can be relaxed by considering the effective change in parameter values. As noted in the Model section, under the strategy of saving a proportion of wealth, the effective interest rate will be $pr$ if a proportion $p$ of wealth is used for gifts, because the donor (individual $i$) will earn $prw_i$ of wealth. If the donor distributes their wealth among $k$ recipients, and if we count the number of gifts in this event as $k$, they will earn $rw_i/k$ in each of the $k$ gifts. Hence, $l$ increases by a factor of $k$, and $r$ decreases by a factor of $1/k$ due to this strategy.
If we count this event as a single gift, then each transaction with each recipient should be valued as contributing $1/k$ of importance, and the donor will earn $rw_i$ of wealth in one gift. The effective parameter values will be $r$ and $l$. Therefore, regardless of how we count this event, the effective value of $rl$ remains unchanged.
Precisely speaking, while this distribution strategy may affect the variance of the noise term in Eq. \eqref{eq:langevin2}, its effect will be negligible unless donors distribute their wealth among $k$ recipients, with $k \sim N$.
Additionally, we assume that any form of wealth transfer equally strengthens the edge weights. Although there may be differences in the strength of contributions in building social connections, considering such differences will result in only minor quantitative changes to phase boundaries. For example, in the current case, creditors earn $1/l$ of score in each step while their debtors are in debt. If we change this rule so that $u/l$ of score is gained, Eq. \eqref{eq:score_langevin} will have the term $u\min(|rw_i - w_j|+, 1)$ instead of $\min(|rw_i - w_j|+, 1)$. This change results only in a quantitative difference.}

In this study, we analyzed a simple model to demonstrate the social changes driven by competitive gift-giving. By simplifying the model, we derived the condition that depends on the scaling parameter $rl$. In our previous model, which considered both kinship and gift-giving, the results also showed the development of socioeconomic disparities as $r$ and $l$ increased.
The phase boundaries depended almost only on $rl$ when $l$ was sufficiently large. However, when $l$ was small, they deviated slightly from the $rl$ scaling due to kinship relationships. Moreover, it is expected that if people sometimes refuse to receive gifts, larger values of $rl$ are needed for the development of disparities. Even in such cases, the degree of deviation from the $rl$ scaling will highlight the specific characteristics of the cases. Other mechanisms, such as warfare and the control of wealth production, will also influence social change \cite{earle1989evolution}.
How such factors cause deviations from the current ideal system must be considered on a case-by-case basis; the present study provides an essential mechanism for the ideal system.

Moreover, even within the same societies, different types of gifts may have different interest rates $r$ and frequencies $l$. Still, the phase diagram serves as a useful reference for discussing specific scenarios. When different types of gifts with different parameters are present, their combined driving forces are expected to generate disparities and drive transitions in social organization. Such scenarios, while beyond the scope of the current model, could effectively correspond to different cases in another parameter region. For example, the sum of $rl$ values might represent the effective $rl$ for such a mixed system.

In conclusion, by proposing a simple model of gift interactions, we demonstrate the emergence of four phases of social organizations characterized by the degrees of economic and social disparities. As the interest rate $r$ and frequency of gifts $l$ increase, wealth distribution shifts from an exponential distribution to a power law distribution, and the score distribution follows. Further, for extremely large values of $r$ and $l$, the score distribution reverts to exponential, accompanied by the emergence of an exceptional outlier, namely, the monarch. In addition, we analytically derive the phase boundaries governed by $rl$ values. The current work explains the basic causal mechanism of social change driven by gift-giving, which expands the scope of econo- and socio-physics and offers a theoretical framework for social sciences from a statistical physics perspective.

\section*{Acknowledgments}
The authors thank Koji Hukushima, Tetsuhiro S. Hatakeyama, and Kim Sneppen for the stimulating discussion.

\end{document}